\documentclass[12pt]{amsart}
\usepackage{hyperref}

\newcommand{\fullref}[1]{\ref{#1} on page~\pageref{#1}}

\newcommand{\ndash}{\nobreakdash-\hspace{0pt}}
\newcommand{\Ndash}{\nobreakdash--}

\newcommand{\ii}{{\mathrm{i}}}
\newcommand{\dd}{{\mathrm{d}}}

\DeclareMathOperator{\Map}{Map}

\newcommand{\id}{\mathrm{id}}

\newtheorem{Thm}{Theorem}[section]

\newtheorem*{Thm*}{Theorem}
\newtheorem*{Lem*}{Lemma}

\theoremstyle{remark}
\newtheorem{Rem}[Thm]{Remark}

\theoremstyle{definition}

\newtheorem{Exa}[Thm]{Example}

\newcommand{\bbR}{{\mathbb{R}}}

\newcommand{\bbZ}{{\mathbb{Z}}}

\newcommand{\de}{\partial}

\newcommand{\calB}{\mathcal{B}}
\newcommand{\calH}{\mathcal{H}}
\newcommand{\calS}{\mathcal{S}}

\newcommand{\calL}{\mathcal{L}}
\newcommand{\calY}{\mathcal{Y}}

\newcommand{\calE}{\mathcal{E}}

\newcommand{\calF}{\mathcal{F}}

\def\gpd{\,\lower1pt\hbox{$\longrightarrow$}\hskip-.24in\raise2pt
               \hbox{$\longrightarrow$}\,}

\newcommand{\EE}{{\mathrm{e}}}

\let\Tilde=\widetilde

\let\Hat=\widehat

\DeclareMathOperator{\im}{im}
\DeclareMathOperator{\coker}{coker}

\newcommand\qq{}
\newcommand\cmp[1]{{\qq Commun.\ Math.\ Phys.\ \bf #1}}
\newcommand\jmp[1]{{\qq J.\ Math.\ Phys.\ \bf #1}}
\newcommand\pl[1]{{\qq Phys.\ Lett.\ \bf #1}}

\newcommand\phr[1]{{\qq Phys.\ Rev.\ \bf #1}}

\newcommand\lmp[1]{{\qq Lett.\ Math.\ Phys.\ \bf #1}}

\newcommand\ijmp[1]{{\qq Int.\ J. Mod.\ Phys.\ \bf #1}}

\begin{document}
\title{On Time}


\author[A.~S.~Cattaneo]{Alberto~S.~Cattaneo}
\address{Institut f\"ur Mathematik, Universit\"at Z\"urich\\
Winterthurerstrasse 190, CH-8057 Z\"urich, Switzerland}  
\email{alberto.cattaneo@math.uzh.ch}
\author[M.~Schiavina]{Michele Schiavina}
\address{Institut f\"ur Mathematik, Universit\"at Z\"urich\\
Winterthurerstrasse 190, CH-8057 Z\"urich, Switzerland} 
\curraddr{Department of Mathematics,
University of California at Berkeley,
887 Evans Hall, 
Berkeley, CA 94720-3840 USA}
\email{michele.schiavina@berkeley.edu}

\thanks{We acknowledge partial support of SNF Grant No.\ 200020-149150/1. 
This research was (partly) supported by the NCCR SwissMAP, funded by the Swiss National Science Foundation, and by the 
COST Action MP1405 QSPACE, supported by COST (European Cooperation in Science and Technology).}




\begin{abstract}
This note describes the restoration of ÔtimeÕ in one-dimensional parameterization-invariant (hence ÔtimelessÕ) models, namely the classically-equivalent Jacobi action and gravity coupled to matter. It also serves as a timely introduction by examples to the classical and quantum BV-BFV formalism as well as to the AKSZ method.
\end{abstract}

\maketitle

\tableofcontents

\section{Introduction}
This note has several purposes. First, it is a timely exercise to explain the classical and quantum BV-BFV formalism of \cite{CMR1,CMR2,CMR3} in simple yet nontrivial one-dimensional models. 

Second, the models we consider---the Jacobi action \cite{J} and the classically equivalent one-dimensional gravity coupled to matter where the cosmological constant plays the role of energy (the one-dimensional versions of
the Nambu--Goto and of the Polyakov actions, respectively)---are interesting as toy models of parametrization invariant actions: the physics they describe has no time. 
We will see however that, not unexpectedly, time may be recovered if one lets  the cosmological constant be a variable (or even a field). In the quantum version we will see that time may also be recovered if one simply allows the initial state to be in a superposition of different cosmological constants. The cosmological constant may also have a  probability distribution peaked at some value so that it will be almost constant.

Third, the study of these simple models requires the whole machinery of the BV-BFV formalism, including a careful discussion of background and residual fields. 
In particular, background independence for 1d gravity with matter can be checked straightforwardly.
The BV Jacobi theory is also an example where the symmetries can be described in bulk equivalent but boundary inequivalent ways. In particular we show that in one version the boundary theory is singular whereas in the other it is not and describes the reduced phase space correctly.

We cannot expect to generalize these results directly to higher dimensional gravity theories, yet the observed phenomena (time independence and possibly singular boundary structure) are a common feature on which these simple models may shed some light. The 4d Palatini--Holst model, treated naively, also yields a singular BV-BFV structure \cite{MS,CS2}  analogous to the first treatment of the Jacobi action presented in this paper
(unlike the Einstein--Hilbert action where everything is regular \cite{MS,CS1}).

The full BV-BFV quantization produces a boundary state which is closed under a specific quantization of the boundary
BFV action. In simpler terms, the bulk theory forces the correct quantization of the constraints. In the one-dimensional examples this just amounts to the Schr\"odinger quantization, but one cannot be misled into thinking that this is a general feature (see \cite{CMR3} for examples where corrections appear, e.g., in the Poisson sigma model). This means that an application of this method to 4d gravity should produce the correct, bulk compatible 
Wheeler--DeWitt equation, but one should in principle expect higher corrections in $\hbar$.

Although some of the results and constructions in this paper are just a rewriting, we think it is worth presenting them in a new light, both
to point out interesting features of the models and to introduce the
BV-BFV formalism using concrete, hands-on examples.

The paper is structured as follows: We start introducing the models and discussing their classical features in Section~\ref{s:classical theory}. 
In Section~\ref{s:redps} we proceed to discuss the underlying symplectic structures following \cite{KT}. 
In Section~\ref{s:BV-BFV formalism} we recall the BV formalism and the construction of the BV-BFV formalism in general as well as applied to these models; this may be viewed as an introduction by examples to the formalisms developed in \cite{CMR1,CMR2}. 
Next, in Section~\ref{s:Quantization}, we discuss quantization of the 1d-gravity-with-matter model; this also serves as an introduction by examples to quantization in the BV-BFV formalism \cite{CMR3} in the presence of a background. Finally, in Section~\ref{s:AKSZ}, we recall the AKSZ construction \cite{AKSZ}, especially in the one-dimensional case: the AKSZ construction yields a topological field theory starting from a target with a certain structure, which in the one-dimensional case is nothing else than an exact BFV structure. It is then not unexpected that we can recover 1d gravity with matter by AKSZ applied to its BFV boundary space derived in Section~\ref{s:BV-BFV formalism}.
We finish discussing some variations on this theme yielding in particular the BV-BFV quantization of 1d gravity with matter in curved target space and its supersymmetric extension, the spinning particle (see also \cite{BDZDVH,G1,G2}).

The BV-BFV formalism for 1d gravity with matter, Section~\ref{s:BFVonedgrav}, and for the minisuperspace formulation of gravity by Hartle and Hawking \cite{HH}, Section~\ref{s:AKSZ generalizations},
have been first discussed in \cite{MS}.

\subsection*{Acknowledgements} We thank the referee for a number of very valuable comments.

\section{The classical theory}\label{s:classical theory}
We start recalling the classical action principle in mechanics. For simplicity we assume the configuration space to be an open
subset $U$ of $\bbR^n$ (even though all the rest of the paper extends to general manifolds). The potential $V$ is a given function on $U$. The kinetic energy $T$ is a quadratic form on $\bbR^n$. For definiteness we take $T(\dot q)= \frac12m||\dot q||^2$, where $||\ ||$ denotes the Euclidean norm. The Lagrangian function is
$L(q,\dot q)=T(\dot q)-V(q)$. The Hamiltonian function $H$, its Legendre transform, then reads
$H(p,q)=\frac{||p||^2}{2m}+V(q)$. The classical action in Hamiltonian form on a fixed interval $[a,b]$ is
\[
S[q]=\int_a^b(p\cdot \dd q-H\dd t)=\int_a^b(p\cdot \dot q-H)\,\dd t
\]
and the equations of motion are obtained as the extremal paths of $S$.

If energy is fixed, $H(p,q)=E$, the orbits are obtained as the extremal paths of the abbreviated action
\[
S_0[q]=\int_a^bp\cdot \dd q=\int_a^bp\cdot \dot q\,\dd t
\]
as one can omit the last term $\int H\dd t = \int E \dd t$ (but we will return on this seemingly innocent passage).
This is Maupertuis' principle.

\subsection{The Jacobi action}
If the Hamiltonian is as above, we have $p\cdot \dot q=2T(\dot q)=2(E-V(q))$. The abbreviated action can then be rewritten as any product $2T^\alpha(E-V)^\beta$ with $\alpha+\beta=1$. The geometric mean,
$\alpha=\beta=\frac12$, yields the Jacobi action \cite{J}
\[
S^J[q]=\int_a^b 2\sqrt{(E-V(q))\,T(\dot q)}\;\dd t.
\] 
This form of the abbreviated action has several cute peculiarities starting with its geometrical interpretation as the length in the target metric
\begin{equation}\label{e:Jacobimetric}
\dd s^2 = 2m(E-V)\dd q^2.
\end{equation}
As such the Jacobi action is parametrization invariant. The orbits are recovered as extremal paths of
fixed energy 
\begin{equation}\label{e:energy}
T(\dot q)+V(q)=E. 
\end{equation}
If we do not fix the energy, the extremal paths---i.e., the geodesics for the
metric \eqref{e:Jacobimetric}---are the classical trajectories but with an arbitrary parametrization: time is lost.
\begin{Rem}[The space of fields]
Because of the special form of the Lagrangian of the Jacobi action, we cannot allow all posible paths. The space of allowed paths, which we will call the space of fields, is
\[
F_{[a,b]}=\{
q\colon[a,b]\to U\ | \ \dot q \not= 0\text{ and } V(q)<E
\}.
\]
Note that, if $E$ is not larger than the minimum of $V$, the space of fields is empty.
\end{Rem}

As remarked above, the correct time parametrization is recovered by the energy condition \eqref{e:energy}. 
The (philosophical) question now is the following. Suppose we are given the Jacobi action tout court, but we do not know its origin from the classical action principle. The energy condition then looks rather arbitrary, just one of the infinitely many ways of fixing the parametrization. For this reason the Jacobi action is often used as a toy model for the problem of time in General Relativity \cite{B}.

One further observation is that we can recover the special parametrization and reconcile the Jacobi with the Hamilton action principle if we do not discard the energy term $\int E\,\dd t$. Namely, consider the extended Jacobi action
\[
\Hat S^J[q;E)=\int_a^b (2\sqrt{(E-V(q))\,T(\dot q)} - E)\;\dd t
\]
which corresponds to the full action at fixed energy, not just the abbreviated one.
By the notation with mismatched parentheses 
we stress that $S^J$ depends on both a path $q$ (hence the square bracket)
 and an energy variable $E$ (hence the round bracket). Extrema w.r.t.\ to $q$ are the same as before, as the added term does not depend on the path. On the other hand
\[
\frac{\de \Hat S^J}{\de E} = \int_a^b \left(\sqrt{\frac T{E-V}}-1\right)\;\dd t.
\]
This the energy condition \eqref{e:energy} on average.


\begin{Rem}\label{r:extJvsmech}
One way of interpreting the extended Jacobi action is that we take all paths at fixed energy $E$, but then let also energy vary freely. For this reason the extended Jacobi action is expected to be classically equivalent to the original Hamilton action. In Section~\ref{s:extendedJacobiclass} we will see that they actually describe the same dynamics
(though for simplicity we will consider only the free particle case $V=0$).
\end{Rem}

Another remark is that, as observed above, the Jacobi action is reparametrization invariant whereas the extended Jacobi action is not, even if by a rather silly path-independent term.
\begin{Rem}\label{r:extendedJacobifields}
A possible generalization consists
in letting $E$ 
depend on time, i.e., become a field. 
The space of fields 
is then
\[
\Hat F_{[a,b]}=\{
q\colon[a,b]\to U,\ E\colon[a,b]\to\bbR
\ | \ \dot q > 0\text{ and } V(q)>E
\}.
\]
We will call this the generalized Jacobi theory.
As $E$ is now a field, criticality is given also by the vanishing of the functional derivative
\[
\frac{\delta \Hat S^J}{\delta E} = \sqrt{\frac T{E-V}}-1,
\]
which is the energy condition \eqref{e:energy}.
\end{Rem}

\subsection{One-dimensional gravity}\label{s:Onedgravuuu}
Another action which is classically equivalent to the Jacobi action is
\begin{equation}\label{e:qg}
S[q,g]:=\int_a^b\left(
\frac1{\sqrt g}T(\dot q) -\sqrt g V(q) + \sqrt g E
\right)\dd t,
\end{equation}
where $g\colon[a,b]\to\bbR_{>0}$ is a new field. In fact we have
\begin{equation}\label{e:deltaSg}
\frac{\delta S}{\delta g} = -\frac12\frac1{g^\frac32}T+\frac12\frac1{g^\frac12}(E-V).
\end{equation}
Hence, extremizing w.r.t.\ $g$ yields
\[
g = \frac T{E-V}.
\]
Inserting this value of $g$ into $S$ yields back $S^J$ as a functional of $q$.
\begin{Rem}
The space of fields for this theory is 
\[
F_{[a,b]}=\{
q\colon[a,b]\to U,\ g\colon[a,b]\to\bbR_{>0}
\}.
\]
\end{Rem}

The action \eqref{e:qg} has the very nice interpretation of one-dimensional gravity coupled to matter. In fact, we may think of $g$ as the component of a source metric on the interval $[a,b]$:
\[
\dd s^2 = g\,\dd t^2.
\]
The term $\sqrt g\,\dd t$ is then just the Riemannian density of this metric, whereas
$g^{-1}\frac12m||\dot q||^2$ is just the kinetic term in the source metric $g$ for the matter field $q$.
Notice, that the Einstein--Hilbert term for gravity is absent in one dimension and that $E$ plays the role of the cosmological constant.
Thus, \eqref{e:qg} is even more to the point a toy model for General Relativity and the problem of time.

Similarly, the extended Jacobi action is classically equivalent to the extended gravity theory
\[
\Hat S[q,g;E):=\int_a^b\left(
\frac1{\sqrt g}T(\dot q) -\sqrt g V(q) + \sqrt g E-E
\right)\dd t,
\]
and the last term breaks again the reparametrization invariance, although in a silly way. 
Again we use mismatched parentheses to stress that $q$ and $g$ are fields whereas $E$ is a variable.
Criticality now also requires the vanishing of
\begin{equation}\label{e:gravdeltag}
\frac{\de\Hat S}{\de E}= \int_a^b(\sqrt g -1)\,\dd t.
\end{equation}
In Section~\ref{s:extendedgravityclass} we will see that the extended gravity action  and the 
original Hamilton action describe the same dynamics.

\begin{Rem}\label{r:extendedgravityfields}
As in the case of the generalized Jacobi action, see Remark~\ref{r:extendedJacobifields}, we can regard $E$ as an additional field. The space of fields 
is then
\[
\Hat F_{[a,b]}=\{
q\colon[a,b]\to U,\ g\colon[a,b]\to\bbR_{>0},\ E\colon[a,b]\to\bbR
\}.
\]
We will call this the generalized one-dimensional gravity.
Criticality with respect to $E$ is now given as the vanishing of the functional derivative
\[
\frac{\delta\Hat S}{\delta E}= \sqrt g -1,
\]
which yields the homogeneous metric condition 
\begin{equation}\label{e:g=1}
g=1. 
\end{equation}
If we insert it into $\Hat S$, we get back 
$\int_a^b (T(\dot q)-V(q))\,\dd t$, i.e.,
the action functional
in the Lagrangian formalism.
\end{Rem}

\subsection{Euler--Lagrange equations}
We claimed above that the Euler--Lagrange (EL) equations for the various theories are all equivalent.
We now justify these claims. 

\subsubsection{The Jacobi action}
We start with the (extended) Jacobi action. At fixed $E$ we have
\[
\delta S^J = \delta \Hat S^J = \int_a^b\sum_i
\left(
\sqrt{\frac{E-V}T}m\dot q^i\delta\dot q^i
-\sqrt{\frac T{E-V}}\,\de_iV\delta q^i
\right)\dd t.
\]
Integrating by parts we get
\begin{multline}\label{e:deltaSJ}
\delta S^J = \delta \Hat S^J =\\= -\int_a^b\sum_i
\left(
\frac\dd{\dd t}\left(
\sqrt{\frac{E-V}T}m\dot q^i
\right)+
\sqrt{\frac T{E-V}}\,\de_iV
\right) \delta q^i\dd t
+\\
+ \sqrt{\frac{E-V}T}m\sum_i\dot q^i\delta q^i\Big|_a^b.
\end{multline}
If we kill the boundary terms, e.g.\ by fixing $q$ at the end points, we get the EL equations
\[
\frac\dd{\dd t}\left(
\sqrt{\frac{E-V}T}m\dot q^i
\right)=
-\sqrt{\frac T{E-V}}\,\de_iV,
\]
which yield back the usual EL equations if we use the energy condition $E-V=T$.
\begin{Rem}
If we consider $E$ as a field, see Remark~\ref{r:extendedJacobifields}, then we get 
\[
\sqrt{\frac T{E-V}}=1
\]
as an additional EL equation. The two EL equations are now equivalent to the usual EL equations in mechanics plus the condition that $E=T+V$. Note that it follows that the field $E$ is constant if it is a solution.
\end{Rem}

\subsubsection{One-dimensional gravity}
We now pass to the (extended) gravity action with matter. We have, at fixed $E$,
\[
\delta S = \delta \Hat S=
\int_a^b \frac{\delta S}{\delta g} \delta g +
\int_a^b\sum_i
\left(
\frac m{\sqrt g}\dot q^i\delta\dot q^i-\sqrt g\,\de_iV\delta q^i
\right)\dd t.
\]
with $\frac{\delta S}{\delta g} $ computed in \eqref{e:deltaSg}. Integrating by parts we get
\begin{multline}\label{e:deltaSgrav}
\delta S = \delta \Hat S=
\int_a^b \frac{\delta S}{\delta g} \delta g -
\int_a^b\sum_i
\left(
\frac\dd{\dd t}\left(\frac m{\sqrt g}\dot q^i\right)
+\sqrt g\,\de_iV
\right)\delta q^i\dd t
+\\
+ \sum_i \frac m{\sqrt g}\dot q^i\delta q^i\Big|_a^b.
\end{multline}
Again, if we kill the boundary terms, e.g., by fixing $q$ at the end points, we get the EL equations
\begin{subequations}
\begin{gather}
g = \frac T{E-V},\label{e:ELg}\\
\frac\dd{\dd t}\left(\frac m{\sqrt g}\dot q^i\right) =
-\sqrt g\,\de_iV.\label{e:ELq}
\end{gather}
\end{subequations}
If we use the homogenous metric
condition \eqref{e:g=1}, then we get the energy condition \eqref{e:energy} and the usual EL equations.
\begin{Rem}
If we consider $E$ as a field, see Remark~\ref{r:extendedgravityfields}, then we get 
$g =1$
as an additional EL equation. The three EL equations are now equivalent to the usual EL equations in mechanics plus the condition that $g=1$ and the condition
$E=T+V$. Note again that it follows that the field $E$ is constant if it is a solution.
\end{Rem}

\section{Symplectic structures and reduced phase spaces}\label{s:redps}
In this section we discuss the boundary symplectic structures induced by the (extended) Jacobi action and by the (extended) one-dimensional gravity. This is a special example of the construction of \cite{KT,CMR2} to which we refer for a more general perspective. In the case of the (extended) Jacobi action we will focus on the simpler case 
$U=\bbR^n$ and $V=0$.
\begin{Rem}\label{r:basicdeclassical}
 We just recall the basic facts, which will be spelled out again in the examples below.
  The boundary term in the variation of the action
can be interpreted as a $1$\ndash form $\Check\alpha$ on the space of fields restricted to the boundary and on their
transversal derivatives.
The space of boundary fields $F^\de$ is then defined by quotienting this space by the kernel of the
 $2$\ndash form $\Check\omega:=\dd\Check\alpha$. This way $F^\de$ is a symplectic manifold, 
 with a symplectic form that we denote by $\omega^\de$,
 and one gets
 a surjective submersion $\pi$ from the space of bulk fields to $F^\de\times F^\de$ by evaluating at the end points
 of the time interval $[a,b]$. The image $L_{[a,b]}$ under $\pi$ of the solutions to the EL equations will represent
 time evolution from time $a$ to time $b$ (in the case of a regular Lagrangian, $L_{[a,b]}$ is just the the graph
 of the Hamiltonian flow). One further complication arises when not every point in $F^\de$ may provide an initial condition. One then defines $C\subset F^\de$ as the set of points that can be completed to a pair in $L_{[a,b]}$ for some
 interval $[a,b]$. The restriction of $\omega$ to $C$ is in general degenerate and one defines the
 reduced phase space $\underline C$ as the quotient of $C$ by its kernel. One can then project $L_{[a,b]}$
 to $\underline C\times\underline C$ and get the correct time 
 evolution $\underline L_{[a,b]}$.\footnote{To be more precise, the $1$\ndash form $\Check\alpha$
 is in general defined on the space $\Check F^\de$ of jets of the fields at a boundary point. The space of boundary
 fields $F^\de$ is then defined as the reduction of $\Check F^\de$ by the kernel of $\Check\omega$ and is assumed
 to be smooth. The relation $L_{[a,b]}$ is then automatically isotropic (i.e., the restriction of $\omega$ to it is zero) and
 is assumed to be Lagrangian (i.e., of maximal dimension) for the theory to be well-defined. The last condition in particular implies that $C$ is coisotropic (i.e., that it is locally defined by constraints whose Poisson bracket vanishes on $C$, a.k.a.\ first class constraints). In this paper all the $F^\de$'s are smooth and all the $L_{[a,b]}$'s are Lagrangian.}
 \end{Rem}

\subsection{The (extended) Jacobi action}\label{s:extendedJacobiclass}
For simplicity we consider only the case $V=0$. This also requires working with $E>0$. 
The boundary term in
\eqref{e:deltaSJ} defines the $1$\ndash form
\[
\Check\alpha = \sqrt{\frac{E}T}m\sum_i v^i\dd q^i =
\sqrt{2mE}\sum_i \frac{v^i}{||v||}\dd q^i, 
\]
where $v$ denotes the velocity. It is better to make a change of variables introducing $\rho:=||v||\in\bbR_{>0}$
and $u:=v/\rho\in S^{n-1}$.
We then have
\[
\Check\alpha = \sqrt{2mE}\sum_i {u^i}\dd q^i. 
\]
In the nonextended case, $E>0$ is just a parameter. Hence,
\[
\Check\omega:=\dd\Check\alpha = \sqrt{2mE}\sum_i {\dd u^i}\dd q^i. 
\]
This $2$\ndash form is clearly degenerate, first because it does not depend on the coordinate $\rho$ and second
because the remaining space $S^{n-1}\times\bbR^n$ is odd dimensional. One can easily verify that its kernel is spanned by the vector fields
\[
X_1 = \frac\de{\de\rho},\qquad X_2= \sum_i u^i\frac\de{\de q^i}
\]
(notice that $\sum_i u^i\dd u^i=0$). If we quotient out by the involutive distribution generated by these vector fields, we end up with the symplectic manifold of boundary fields
$F^\de = TS^{n-1}$, where the symplectic structure is given by $\sqrt{2mE}$ times the pull back of the canonical
symplectic structure on $T^*S^{n-1}$ by the identification $TS^{n-1}\to T^*S^{n-1}$ given by the restriction of the Euclidean metric. Notice that the $u^i$s are the coordinates on the base $S^{n-1}$. The fiber coordinates $[q^i]$ are given by the coordinates $q^i$ modulo translations by $u^i$. We have a surjective submersion
\[
\pi\colon\begin{array}[t]{ccc}
F_{[a,b]} &\to &F^\de\times F^\de\\
q &\mapsto &\left(\frac{\dot q(a)}{||\dot q(a)||},[q(a)];
\frac{\dot q(b)}{||\dot q(b)||},[q(b)]
\right)
\end{array}
\]
The EL equations reduce, for $V=0$, to
\[
\frac\dd{\dd t} \frac{\dot q}{||\dot q||}=0.
\]
We define $EL_{[a,b]}\subset F_{[a,b]}$ as the space of solutions and set 
\[
L_{[a,b]}:=\pi(EL_{[a,b]})\subset F^\de\times F^\de.
\]
In a solution the quantity $u:=\dot q/||\dot q||$ is constant. Moreover, we have
\[
q^i(b)-q^i(a) = u^i\int_a^b ||\dot q(t)||\dd t.
\]
This implies that $L_{[a,b]}$ is the graph of the identity map. 
Hence this result says that, as expected, there is no time evolution.

We now pass to the extended Jacobi action. The construction of the boundary symplectic
space proceeds as above just by taking into account that now also the variable $E$ belongs to
the space of fields restricted to the boundary.
It is convenient to make one further change of variables
defining $p_i:=\sqrt{2mE}u^i\in\bbR^n\setminus\{0\}$. We then have
$\Check\alpha =\sum_i p_i\dd q^i$ and $\Check\omega:=\dd\Check\alpha = \sum_i \dd p_i\dd q^i$.
Notice that this $2$\ndash form is still degenerate since it does not depend on $\rho$. The kernel is however just
the span of $X_1$ above. The quotient is the symplectic manifold of boundary fields 
$F^\de=\bbR^n\times(\bbR^n\setminus\{0\})$---i.e., $T^*\bbR^n$ minus the zero section---with its canonical
symplectic structure. 
We have 
the surjective submersion
\[
\pi\colon\begin{array}[t]{ccc}
F_{[a,b]} \times \bbR_{>0}&\to &F^\de\times F^\de\\
(q,E) &\mapsto &\left(\frac{\sqrt{2mE}\,\dot q(a)}{||\dot q(a)||},q(a);
\frac{\sqrt{2mE}\,\dot q(b)}{||\dot q(b)||},q(b)
\right)
\end{array}
\]
The EL equations can be rewritten as
\begin{gather*}
\frac\dd{\dd t}\left(\sqrt{2mE} \frac{\dot q}{||\dot q||}\right) = 0,\\
\sqrt\frac m{2E}\int_a^b ||\dot q(t)||\,\dd t = b-a.
\end{gather*}
The first equation implies that $P:=\sqrt{2mE}\dot q/{||\dot q||}$ is constant, so that the initial and the final momenta
in $F^\de$ are equal. It also implies that
\[
q(b)-q(a)=\frac P{\sqrt{2mE}}\int_a^b ||\dot q(t)||\,\dd t=\frac Pm\,(b-a),
\]
where we have also used the second EL equation. Hence, if we define $L_{[a,b]}$ as $\pi(EL_{[a,b]})$ with $EL_{[a,b]}$ the
space of solutions to the EL equations, we get
\begin{equation}\label{e:LabextJ}
L_{[a,b]} = \left\{(q,p;\Tilde q,\Tilde p)\in F^\de\times F^\de\ | \ \Tilde p = p,\ \Tilde q =q+\frac pm (b-a)
\right\}.
\end{equation}
Thus, we recover the usual time evolution for the free particle, as promised in Remark~\ref{r:extJvsmech}.
\begin{Rem}
In the case of the  generalized Jacobi theory of Remark~\ref{r:extendedJacobifields},
where $E$ is a field, the boundary symplectic manifold
$F^\de$ is the same as in the extended theory but now we have 
the surjective submersion
\[
\pi\colon\begin{array}[t]{ccc}
\Hat F_{[a,b]} &\to &F^\de\times F^\de\\
(q,E) &\mapsto &\left(\frac{\sqrt{2mE(a)}\,\dot q(a)}{||\dot q(a)||},q(a);
\frac{\sqrt{2mE(b)}\,\dot q(b)}{||\dot q(b)||},q(b)
\right)
\end{array}
\]
The EL equations read
\begin{align*}
\frac\dd{\dd t}\left( \frac{\sqrt{2mE}\dot q}{||\dot q||} \right)&= 0,\\
\frac12 m||\dot q||^2 &= E.
\end{align*}
Following the same reasoning as above we conclude again that $L_{[a,b]}$
is as in \eqref{e:LabextJ}.
%
\end{Rem}

\subsection{The (extended) one-dimensional gravity}\label{s:extendedgravityclass}
We start with the nonextended model, where the fields are $q$ and $g$, whereas $E$ is a parameter.
{}From \eqref{e:deltaSgrav} we get the $1$\ndash form
\[
\Check\alpha = \sum_i \frac{m \dot q^i}{\sqrt g}\,\dd q^i.
\]
If we introduce $p_i:=m \dot q^i/\sqrt g$, we see that the space of boundary fields $F^\de$ is just
$T^*U$ with canonical symplectic form $\omega=\sum_i \dd p_i\dd q^i$. 

In the nonextended theory we then have the surjective submersion
\begin{equation}\label{e:pigrav}
\pi\colon\begin{array}[t]{ccc}
F_{[a,b]} &\to &F^\de\times F^\de\\
(q,g) &\mapsto &\left(\frac{m \dot q^i(a)}{\sqrt{g(a)}},q(a);
\frac{m \dot q^i(b)}{\sqrt{g(b)}},q(b)
\right)
\end{array}
\end{equation}
If we define $p_i(t):=m \dot q^i(t)/\sqrt{g(t)}$, then the first EL equation \eqref{e:ELg} yields
\begin{equation}\label{e:T+V=E}
\frac{||p||^2}{2m}+V(q) = E,
\end{equation}
which implies that the space $EL_{[a,b]}$ 
of solutions to the EL equations is empty unless $E$ is greater or equal to the minimum of $V$. To avoid singularities,
we are going to assume next that $E$ is strictly greater than the minimum.
The second EL equation \eqref{e:ELq} together with the definition of $p$ may be rewritten as the system
\begin{equation}\label{e:ELgravHamilton}
\begin{split}
\frac1{\sqrt g}\,\dot p^i &= -\de_i V,\\
\frac1{\sqrt g}\,\dot q^i &= \frac{p_i}m.
\end{split}
\end{equation}
If we define $\tau$ as the solution to $\frac{\dd\tau}{\dd t}=\sqrt g$, $\tau(a)=0$, we then get Hamilton's equations
\begin{align*}
\frac{\dd p^i}{\dd\tau} &= -\de_i V,\\
\frac{\dd q^i}{\dd\tau} &= \frac{p_i}m.
\end{align*}
We then conclude that $L_{[a,b]}:=\pi(EL_{[a,b]})$, where $EL_{[a,b]}$ is the space of solutions to the EL equations, is
\begin{multline*}
L_{[a,b]} = \Big\{(p,q;\Tilde p,\Tilde q)\in T^*U\times T^*U\ \Big|\ 
\frac{||p||^2}{2m}+V(q) = E \\ \text{ and }\exists s>0 : (\Tilde p,\Tilde q)=\Phi_s^H(p,q)
\Big\}
\end{multline*}
where $\Phi_s^H$ denotes the flow at time $s$ of the Hamiltonian vector field of the Hamiltonian
$H(p,q)=T(p)+V(q)$.\footnote{Note that $L_{[a,b]}$ is not a graph since on the source $T^*U$
one has to impose the constraint $\frac{||p||^2}{2m}+V(q) = E$. On the other hand, since the flow is
for a nonfixed time $s$, we have that anyway $\dim L_{[a,b]}=2n$ and $L_{[a,b]}$ is indeed Lagrangian.}
Because of the constraint $\frac{||p||^2}{2m}+V(q) = E$ not every point in $F^\de=T^*U$ is a possible initial condition.
Indeed the space of possible initial conditions is
\begin{equation}\label{e:cleanest}
C = \left\{(p,q)\in T^*U\ \Big|\ \frac{||p||^2}{2m}+V(q) = E\right\}.
\end{equation}
Its reduction $\underline C$ by the kernel of the restriction of $\omega$ consists of equivalence classes
where $(p,q)\sim(\Tilde p,\Tilde q)$ if $(\Tilde p,\Tilde q)=\Phi_s^H(p,q)$ for some $s$. As a consequence, the reduction
$\underline L_{[a,b]}$ turns out to be, as expected, the graph of the identity.
\begin{Rem}
The reduced phase space $\underline C$ turns out to be the same as for the Jacobi theory.
We check this for $U=\bbR^n$ and $V=0$. In this case, $C$  
from equation \eqref{e:cleanest}
is an $n$\ndash dimensional sphere, of radius $\sqrt{2mE}$, times $\bbR^n$ and the reduction is by the vector field
$\sum_i p_i\frac\de{\de q^i}$.
\end{Rem}

We now turn to the extended theory. Recall that now also the variable $E$ belongs to
the space of fields restricted to the boundary. Since, however, it does not enter the $1$\ndash form $\Check\alpha$,
it is modded out when we pass to the quotient $F^\de$.
The space of boundary fields $F^\de$ is then again $T^*U$ with canonical symplectic structure. The surjective submersion $\pi\colon F_{[a,b]}\times\bbR\to  F^\de\times F^\de$ is given
by the same map as in \eqref{e:pigrav} since it does not depend on $E$. We now compute $L_{[a,b]}$.
The first difference is that now $E$ is a free parameter, so equation \eqref{e:T+V=E} is not a constraint on $p$ and $q$, but just fixes $E$. The second difference is that we now also have the EL equation
\[
\int_a^b\sqrt g\,\dd t = b-a,
\]
which follows from \eqref{e:gravdeltag}. This simply implies that $\tau(b)=b-a$. Hence, 
\begin{equation}\label{e:Lextgrav}
L_{[a,b]}=\{ F^\de\times F^\de\ |\ 
(\Tilde p,\Tilde q)=\Phi_{b-a}^H(p,q)
\},
\end{equation}
that is, the graph of the Hamiltonian flow, as promised in Section~\ref{s:Onedgravuuu}.
\begin{Rem}
In the generalized theory of Remark~\ref{r:extendedgravityfields}, the analysis is not that different.
First, $F^\de$ is still $T^*U$ with canonical symplectic structure. The surjective submersion $\pi\colon \Hat F_{[a,b]}\to  F^\de\times F^\de$ is still given
by the same map as in \eqref{e:pigrav} since it does not depend on $E$. Again, the first EL equation is not a constraint but just fixes $E$. Finally, the new EL equation $g=1$ simply turns the system \eqref{e:ELgravHamilton} into the usual
Hamilton equations. Hence, $L_{[a,b]}$ is again given by \eqref{e:Lextgrav}.
\end{Rem}

\subsection{Conclusions}\label{s:conclusionSymplecticStructure}
In this section we have analyzed the boundary symplectic structures for the Jacobi theory and for 1d gravity with matter. Recall that in the nonextended theories $E$ is a fixed parameter. In the extended theories we add the term
$-\int E\,\dd t$ to the action and treat $E$ as a variable. In the generalized theories we 
add the term $-\int E\,\dd t$ and 
treat $E$ as a field.

The classical analysis shows that the Jacobi theory and 1d gravity coupled to matter are 
equivalent.\footnote{The attentive reader might have noticed that the Jacobi theory actually has a smaller 
phase space as one has to remove values corresponding to singularities of the Lagrangian function (velocity
equal to zero in the example $V=0$). This shows that one-dimensional gravity is an improved version of the Jacobi
theory that cures the singularities.}
In the nonextended versions the reduced phase spaces have dimension $2n-2$ and the evolution is given by the
identity map. The extended and the generalized theories turn out to be equivalent to each other and also to the
usual Hamiltonian theory with phase space $T^*U$ and Hamilton 
evolution.

\section{Symmetries}\label{s:BV-BFV formalism}
Our next goal is to quantize the theories described in the previous section. We plan to do it in the perturbative framework. For this we would need the Hessian of the action around a critical point to be non degenerate. This happens in the generalized theories where the energy becomes a field. In the other cases the way out is the 
BRST formalism \cite{BRS,T} or even better the BV formalism \cite{BV81}, which is more suitable to deal with the boundary, see \cite{CMR1,CMR2}. We start with a general overview of the basic facts that will be spelled out again in the examples below.

\subsection{The BV-BFV formalism}
The first step is to identify the symmetries of the theories. They are just the reparametrizations and at the infinitesimal level they are parametrized by vector fields on the source interval $[a,b]$, so we introduce a new field $\xi$, the ghost, which is a Grassmann variable taking values in vector fields. 
It is convenient for bookkeeping to introduce a $\bbZ$\ndash grading, the ghost number, which is $0$ for the classical fields and $1$ for the ghosts. We then introduce the BRST operator $Q$, of degree $1$, that acts on fields by the Lie derivative. Hence, we have
\[
Qq=\xi\dot q
\]
and, in the gravity theory,
\[
Qg=\xi\dot g+2g\dot \xi
\]
(recall that $g$ is the component of a rank-$2$ symmetric contravariant tensor field). In the extended theories,
where $E$ is also a variable, we set
\[
QE=0.
\]
The BRST operator is next extended to the ghost itself to encode the Lie bracket of vector fields:
\[
Q\xi=\xi\dot\xi.
\]
As a result $Q^2=\frac12[Q,Q]=0$. One then usually says that $Q$ is a \textbf{cohomological vector field}. Parametrization invariance implies that Q of the action is zero up to boundary terms.

The BV formalism extends this by introducing the so-called antifields: namely, for each field (including the ghosts) one introduces a new field of dual nature (its momentum) with 
opposite Grassmann parity (i.e., odd for the physical fields and even for the ghosts). We call the space of fields and antifields 
the BV
space of fields.
The ghost number of the antifields for the fields is set to $-1$, whereas the ghost number of the antifield for the ghost is set to $-2$. As a result, on the BV space of fields one has a canonical symplectic structure that is odd and has ghost number $-1$, to which we will refer as the BV form.
Next one adds to the original action the sum of the pairings between each antifield and Q of the corresponding field, obtaining the BV
action of the theory.\footnote{The BV action in the present example is linear in the antifields, which is not a general
feature of the formalism.} One finally extends the BRST operator to the whole BV space of fields as the Hamiltonian vector field of the extended action, up to boundary terms. One can easily check that this extension, which we keep denoting by $Q$, is still cohomological.
\begin{Rem}
In a more abstract language, the BV space of fields is the cotangent bundle shifted by $-1$ of the space of fields and ghosts with its canonical symplectic structure, and the added term to the action is the Hamiltonian function for the canonical lift of $Q$. The classical action changes the canonical lift by a vertical part.
\end{Rem} 

In the absence of a boundary---i.e., if we work on a circle instead of the interval---we may define the Poisson bracket of the extended action with itself and this turns out to be zero as a consequence of the cohomological nature of 
the BRST operator and of the invariance of the classical action. This is called the \textbf{classical master equation}.

In the presence of a boundary, things may be remedied 
 by the formalism introduced in \cite{CMR1,CMR2}. Namely,
one applies to the BV action the construction described at the beginning of Section~\ref{s:redps}, see Remark~\fullref{r:basicdeclassical} for more details. This produces a symplectic space 
$(\calF^\de,\omega^\de=\delta\alpha^\de)$	
associated to the boundary. One can show that $Q$ projects to a boundary cohomological vector field $Q^\de$ 
(known as the BFV operator \cite{BFV1,BFV2} or the BRS operator \cite{KS}). 
The failure of the classical master equation in the presence of a boundary is now under control, as explained in \cite{CMR1,CMR2}, by the equation
\[
\iota_Q\omega = \delta\calS + \pi^*\alpha^\de,
\]
where $\pi$ is the projection $\calF\to\calF^\de\times\calF^\de$ obtained by restricting the fields to the boundary
(which in the present one-dimensional case consists of two points). It also turns out that $Q^\de$
is Hamiltonian with respect to $\omega^\de$; namely, there is a uniquely defined odd function $\calS^\de$ (the boundary
action) of degree $+1$ satisfying $\iota_{Q^\de}\omega^\de = \delta\calS^\de$.

This construction however requires some regularity assumptions that turn out to fail in the case of the Jacobi action. 
One way out, as we will explain, consists in defining the space of fields in a more appropriate way. Another consists
in redefining the symmetries in a completely equivalent way as long as there is no boundary, but resolving the problems at the boundary. One-dimensional gravity with matter turns out to work well.


\subsection{The (extended) Jacobi action}
Again we only consider the case $V=0$. We start with the non extended case. A simple computation shows that
\[
Q\sqrt T = \frac\dd{\dd t}\left(\xi\sqrt T\right),
\]
which proves the invariance of the action up to boundary terms. To the field $q$ and the ghost $\xi$ we associate
the antifields $q^+$ (odd and of ghost number $-1$) and $\xi^+$ (even and of ghost number $-2$). The symplectic
structure on the BV space of fields reads
\[
\omega_\text{BV} = \int_a^b \left(\sum_i\delta q^+_i\delta q^i + \delta\xi^+\delta\xi\right)\,\dd t.
\]
Here $\delta$ denotes the ``de~Rham differential" on the BV space of fields.
The BV action is
\[
\calS^J[q,q^+,\xi,\xi^+]=
\int_a^b\left(
2\sqrt{ET} + \sum_iq^+_i\xi\dot q^i - \xi^+\xi\dot\xi\right)\,\dd t.
\]
The boundary term in $\delta \calS^J$ yields
\[
\Check\alpha = 
\sum_i (\sqrt{2mE}\,u^i+q^+_i\xi)\dd q^i + \xi^+\xi\dd\xi, 
\]
where $u:=v/||v||$ is the normalized velocity. This yields the boundary $2$\ndash form
\[
\Check\omega = \delta \Check\alpha =
\sum_i(
\sqrt{2mE}\,\dd u^i\dd q^i+\xi\dd q^+_i\dd q^i - q^+_i\dd\xi\dd q^i)+
\xi^+\dd\xi\dd\xi-\xi \dd\xi^+\dd\xi
.
\]
We now claim that the kernel of $\Check\omega$ is singular. Indeed, let
\[
K = \sum_i\left(
X^i\frac\de{\de q^i} + U^i\frac\de{\de u^i} + X^+_i\frac\de{\de q^+_i}
\right)
+\Xi\frac\de{\de\xi} + \Xi^+\frac\de{\de\xi^+}
\]
be a generic (even) vector field. Notice that the normalization constraint $\sum_i u^iu^i=1$
implies $\sum_i U^iu^i=0$ and $\sum_i u^i\dd u^i=0$. The condition that $K$ be in the kernel of $\Check\omega$
yields the following equations:
\begin{gather*}
\sqrt{2mE}\,U^i-\xi X^+_i+q^+_i\Xi=0,\\
\sum_i X^i\dd u^i=0,\\
\xi X^i=0,\\
\sum_iq^+_iX_i
+2\xi^+\Xi+\xi\Xi^+=0,\\
\xi\Xi=0.
\end{gather*}
The first equation may be solved for $U$,
the second equation implies that $X$ is proportional to $u$ (like in the classical case), but the third equation is singular. Notice that for $\xi=0$ (which is a supersubmanifold) there is no condition. On the other hand, we cannot assume that $X$ is proportional to $\xi$ as this would not produce the correct kernel for $\xi=0$. We cannot even put the condition $\xi\not=0$ on the space of fields, for this does not produce a supersubmanifold.\footnote{As
$\xi^+$ is even, we could however impose $\xi^+\not=0$ to make the fourth equation nonsingular and solvable
for $\Xi$. The fifth equation would then be automatically satisfied.}

The way out is to define the base of bulk fields by adding a closed conditions. One choice consists in setting
$\xi=0$ on the boundary. This may look as the natural choice as $\xi$ describes infinitesimal reparametrizations of the interval: from a geometrical viewpoint one might have expected that 
we should have required it from the very beginning to vanish at the boundary. Note however that $\xi$ has indeed a geometric origin, yet once it is introduced in the formalism it is just a field like the others. Moreover, in the other examples
in this paper as well as in the Einstein--Hilbert formulation of general relativity \cite{CS1}
 setting $\xi$ to zero on the boundary would be wrong as it would produce the wrong
reduced phase space.

Another choice instead consists in setting
$q^+=0$ and $\xi^+=0$ on the boundary. Both conditions imply that the boundary space $\calF^\de$ is the same as the classical one: $TS^{n-1}$. 

We finish by remarking that $Q$ is projectable to the boundary, actually to the zero vector field. We have already computed $Qq$ and $Q\xi$.
Deriving by $t$ we get $Q\dot q=\dot\xi\dot q+\xi \ddot q$ which implies $Qu=\xi a_\perp$,
where $a=\frac{\ddot q}{||\dot q||}$ and $a_\perp:= a -(u\cdot a)u=\dot u$. From the BV action we also get
\begin{gather*}
Qq^+ = \frac\dd{\dd t}\left(
\sqrt{2mE}\frac{\dot q}{||\dot q||}+q^+\xi
\right),\\
Q\xi^+=q^+\dot q + \xi\dot\xi^++2\xi^+\dot\xi.
\end{gather*}
If we set $\xi=0$, then $Qq$ and $Qu$ vanish on the boundary. If on the other hand we set $q^+=0$, by consistency we also have to set $Qq^+=0$ on the boundary, i.e., $\sqrt{2mE}a_\perp+\dot q^+\xi=0$. This implies $\xi a_\perp=0$
on the boundary, so that $Qu=0$. Note that $Qq$ is proportional to $u$ and hence produces a vertical term.
In both cases, we end up with $\calF^\de$ being $T^*S^{n-1}$ as in the classical situation.

We now move on to the extended theory. In the bulk the only difference is the addition of the term
$\dd E^+\dd E$ to the BV form. This yields an additional term to $Q$; namely,
\[
QE^+=\int_a^b \sqrt{\frac TE}\;\dd t.
\]
Now $Q^2E^+=\frac1{\sqrt E}\int_a^b \frac\dd{\dd t}(\xi\sqrt{T})\;\dd t$.
If we have a boundary, we must then impose $\xi=0$
there to ensure that $Q$ squares to zero. This implies $\calF^\de=T^*S^{n-1}$ as in the classical situation.

\subsubsection{Rewriting the symmetries}\label{s:uffa}
The natural way to let vector fields act is by Lie derivative as we did at the beginning of this section. As we have seen however, this creates some problems. We can instead let vector fields act on coordinates by
the Lie derivative rescaled by $||\dot q||$. Namely, denoting the vector field ghost by $\zeta$, we set
\[
Qq=\zeta u
\]
with $u = \frac{\dot q}{||\dot q||}$.

A simple computation yields $Q||\dot q||=\dot\zeta$. This implies $Q(2\sqrt{ET})=\sqrt{2mE}\,\dot\zeta$ which proves the invariance of the classical action up to boundary terms.

Another simple computation then yields $Qu=\frac{\zeta\dot u}{||\dot q||}$; hence $Q^2q= (Q\zeta)u$. The only way to make $Q^2$ vanish is then to set
\[
Q\zeta = 0.
\]
Note that the symmetries are now abelian.
We now get the 
BV action
\[
\calS^J[q,q^+,\zeta,\zeta^+]=
\int_a^b\left(
2\sqrt{ET} + \sum_iq^+_i\zeta u^i\right)\,\dd t.
\]
The boundary term in $\delta \calS^J$ now yields
\[
\Check\alpha = 
\sqrt{2mE}\sum_i u^i\dd q^i 
\]
with
\[
\underline u = u + \frac{q^+_\perp\xi}{\sqrt{2mE}||\dot q||}
\]
and $q^+_\perp=q^+-u(u\cdot q^+)$. Notice that $\underline u\cdot \underline u=1$, so $\underline u\cdot\dd\underline u=0$. 
This means that the vector field $X=\underline u\cdot\frac\de{\de q}$ is in the kernel of $\Check\omega=\dd\Check\alpha$ and one can verify that its span is the whole kernel.

As a result, we can identify $\calF^\de$ with $T^*S^{n-1}$ with coordinates $\underline u$ on the base, $\underline p = q\mod \underline u$ on the fiber and
symplectic form $\sqrt{2mE}\,\dd \underline p\cdot\dd \underline q$.
As expected $Q^\de=0$.

\begin{Rem}
If there is no boundary, the two versions of the theory are BV equivalent. The passage from the the ``old'' coordinates $(q,q^+,\xi,\xi^+)$ to the
``new'' ones $q,q^+,\zeta,\zeta^+$ is given by the symplectomorphism
\begin{align*}
\Tilde q &= q, & \Tilde q^+ &= q^+-\frac\dd{\dd t}\left(\frac{\xi^+\xi\dot q}{||\dot q||^2}
\right)
,\\
\zeta &= ||\dot q||\xi, & \zeta^+ &=\frac{\xi^+}{||\dot q||}.
\end{align*}
One can easily verify that this symplectomorphism maps the BV action in one set of coordinates to the BV action in
the other. Note that the transformation depends on derivatives of the fields. It turns then out that in presence of a boundary it is no longer a symplectomorphism. This explains why the two bulk equivalent theories
yield different boundary structures, one where the boundary two-form has a kernel of constant rank
and one where this is not the case.
\end{Rem}

In the extended theory, we have to impose $\xi=0$ on the boundary to ensure that $Q^2=0$. So we get
$\calF^\de=T^*\bbR^n$ with canonical one-form $p\cdot\dd q$ and
$p = \sqrt{2mE}\,u$. 

\subsection{The (extended) one-dimensional gravity}\label{s:BFVonedgrav}
In this case we also have the field $g$ and its antifield $g^+$ (odd of ghost number $-1$). 
The symplectic
structure on the BV space of fields reads
\[
\omega_\text{BV} = \int_a^b \left(\sum_i\delta q^+_i\delta q^i + \delta g^+\delta g +
\delta\xi^+\delta\xi\right)\,\dd t.
\]
The BV action is
\[
\calS[q,q^+\!,g,g^+\!,\xi,\xi^+]=
S[q,g]+
\int_a^b\left( 
\sum_iq^+_i\xi\dot q^i +g^+(\xi\dot g+2g\dot\xi)
- \xi^+\xi\dot\xi\right)\,\dd t.
\]
The boundary term in $\delta \calS$ yields
\[
\Check\alpha = 
\left(\frac{m\dot q}{\sqrt g}+q^+\xi\right)\cdot\dd q+
g^+\xi\dd g + (\xi^+\xi-2g^+g)\dd\xi.
\]
One can then see that $\calF^\de$ can be identified with $T^*(\bbR^n\times\bbR[1])$ with base
coordinates $q,c$ and fiber coordinates $p,b$ and canonical $1$\ndash form
$\alpha^\de = p\cdot\dd q+b\,\dd c$. The projection map is defined by
\begin{equation}\label{e:gravityprojection}
\begin{split}
p &= \frac{m\dot q}{\sqrt g} + q^+\xi,\\
b &= \frac1{\sqrt g}(\xi^+\xi-2g^+g),\\
c &= \sqrt g\,\xi.
\end{split}
\end{equation}

We now have to compute $Q^\de$ and its Hamiltonian function $S^\de$. To simplify computations we use a little trick
(due to Roytenberg \cite{Roy}).
Let
\[
\Check S := \iota_Q\iota_E\Check\omega,
\]
where $E=\xi\frac\de{\de\xi}-2\xi^+\frac\de{\de\xi^+}-g^+\frac\de{\de g^+}-\sum_iq^+_i\frac\de{\de q^+_i}$
is the graded Euler vector field. Since $L_E\beta=|\beta|\beta$ for a differential form of ghost number $|\beta|$,
we get $\dd\Check S=\iota_Q\Check\omega$. Since $Q$ is projectable, this formula says that $\Check S$
is the pullback of a function $S^\de$ and that this function is the Hamiltonian function of $Q^\de$. A simple computation
yields
\[
\iota_E\Check\omega = -(\xi^+\xi-2g^+g)\dd\xi+\dd (\xi^+\xi-2g^+g)\,\xi.
\]
After some simplifications we obtain
\[
\iota_Q\iota_E\Check\omega = -2\,(Qg^+)\,g\xi.
\]
Since
\[
Qg^+ = \frac12\frac1{\sqrt g}\left(E-\frac{T(\dot q)}g-V(q)\right),
\]
we get 
\[
\Check S = \left(\frac{T(\dot q)}g+V(q)-E\right)\sqrt g\xi,
\]
which is the pullback of
\[
S^\de = \left(\frac{||p||^2}{2m}+V(q)-E\right)c.
\]
This is the BFV action to impose the energy constraint, as expected. What is important, for quantization, is
however the projection map \eqref{e:gravityprojection} from bulk to boundary coordinates.
\begin{Rem}[Rewriting the symmetries]\label{r:resymmgrav}
The projection map suggests looking for a BV symplectomorphism between coordinates
$(q,g,\xi,q^+\!,g^+\!,\xi^+)$ and $(\Tilde q,\Tilde g,c,\Tilde q^+\!,\Tilde g^+\!,c^+)$ such that
$c=\sqrt g\,\xi$. Indeed one has such a symplectomorphism (also when there is a boundary) in the form
\begin{align*}
q &= \Tilde q, & q^+ &= \Tilde q^+,\\
g &= \Tilde g & g^+&=\Tilde g^++\frac{c^+c}{2\Tilde g},\\
\xi &= \frac c{\sqrt g} & \xi^+ &= \sqrt g\,c^+.
\end{align*}
This transformation maps the old BV action into
\[
\calS[\Tilde q,\Tilde g,c,\Tilde q^+\!,\Tilde g^+\!,c^+]= S[\Tilde q,\Tilde g] +
\int_a^b\left( 
\sum_i\Tilde q^+_i \frac{c\dot{\Tilde q^i}}{\sqrt{\Tilde g}} 
+2\Tilde g^+ \sqrt{\Tilde g}\,\dot c
\right)\,\dd t.
\]
Notice that this transformation ``abelianizes'' the symmetries as we get $Qc=0$.
Another useful remark is that in the new variables we have
\[
b = -2\Tilde g^+\sqrt{\Tilde g}.
\]
\end{Rem}

We now move on to the extended theory. In the bulk the only difference is the addition of the term
$\dd E^+\dd E$ to the BV form. This yields an additional term to $Q$; namely,
\[
QE^+=\int_a^b\sqrt g\;\dd t.
\]
Now $Q^2E^+= \int_a^b\frac\dd{\dd t}(\xi \sqrt g)\;\dd t$. If we have a boundary, we must then impose $\xi=0$
there to ensure that $Q$ squares to zero. As a result, $\calF^\de$ is $T^*\bbR^n$ with 
$p = \frac{m\dot q}{\sqrt g}$, and of course we have $Q^\de=0$ and $S^\de=0$.

\section{Quantization}\label{s:Quantization}
Theories with symmetries have to be gauge fixed. In the BV formalism gauge fixing corresponds to choosing a Lagrangian
submanifold $\calL$ of the BV space of fields $\calF$ on which the action is no longer degenerate. Integration on a 
Lagrangian submanifold is usually referred to as BV integration.

The quantum master equation $(\calS,\calS)-2\ii\hbar\Delta\calS=0$ ensures invariance under deformations of $\calL$.
Here $\Delta$ denotes the BV Laplacian, a second order differential operator
which is ill defined on an infinite dimensional space. To make sense of $\Delta$ in field theory, one has first to regularize. 

In Darboux coordinates, $\Delta$ is the sum/integral (with signs) of the the second derivatives
with respect to a field and the corresponding antifield. Note that the BV actions we consider are formally $\Delta$\ndash closed
if we use the versions of Section~\ref{s:uffa} and Remark~\ref{r:resymmgrav} as whenever a field and its antifield appear
in the same term one is derived with respect to $t$. In the other versions of the theories we always get a singular term
$\delta(0)\dot\xi$. There may be regularizations in which this term vanishes.

If there is a boundary, quantization is supposed to produce a state, i.e., an element of a vector/Hilbert space
associated to the boundary. If, as in all our examples,  $\calF^\de$ is a cotangent bundle $T^*\calB$
with $\omega^\de$ its canonical symplectic form, we can define the graded vector space $\calH$ to be a space of functions on $\calB$.
The function $\calS^\de$ should then be quantized to an operator $\Omega$ on $\calH$. If $\alpha^\de$ is the canonical one-form of $T^*\calB$, we may use the Schr\"odinger quantization where the coordinates
 on $\calB$ are quantized as multiplication operators, the fiber coordinates (momenta) are quantized as derivatives
 of the corresponding base coordinate (times $-\ii\hbar$), and we put all derivatives to the right (standard ordering).
 The condition $(\calS^\de,\calS^\de)=0$ implies that $\Omega^2=O(\hbar)$. The good situation is however
 when $\Omega^2$ is exactly zero as in this case $(\calH,\Omega)$ is a complex and we may interpret its
 cohomology in degree zero as the physical space (a quantization of the reduced phase space if smooth).
 
 One of the main results of \cite[Section 2.4]{CMR3} is the following: under the assumptions that 
$i)$ the fibration
 $\calF\to\calB$ splits as a product $\calF\simeq\calY\times\calB$ with the BV form 
 on $\calF$
 being the pullback of
 a BV form
 on the factor $\calY$
 and 
 $ii)$ the BV action is $\Delta$\ndash closed (for a formally defined $\Delta$\ndash operator on $\calY$), the state $\psi$ produced by BV integration satisfies the
 modified quantum master equation
 \[
 (\hbar^2\Delta+\Omega)\psi=0,
 \]
 where $\Omega$ is the Schr\"odinger quantization of $\calS^\de$ and now $\Delta$ is the BV Laplacian
 on residual fields.\footnote{To be precise, the result is proved for
 finite dimensional $\calF$. Note that being a pullback the BV form on $\calF$ is necessarily degenerate, unless $\calB$
 is zero-dimensional; it is assumed anyway that that the BV form on $\calY$ is nondenegerate.
 In the infinite dimensional case, however, it is possible (and it turns out to be true in examples) that
 both BV forms, on $\calF$ and $\calY$, are weakly nondegenerate. The general proof of the modified quantum
 master equation
 however becomes formal (note that, to start with, $\Delta$ is not even well defined). 
 The correct viewpoint is that one expects the modified quantum master equation to hold, but one has to prove it explicitly from the perturbation theory.}   
 The latter form a finite dimensional BV space on which we have not performed the BV integration yet. A typical example is the Wilson approach to renormalization, where the residual fields are the infrared components
 in the splitting of the fields into infrared and ultraviolet. Another situation is when we take the residual fields
 to be classical solutions, modulo symmetries, of the Euler--Lagrange equations (zero modes). 
Fixing a space of residual fields is also important for the following two reasons:
 firstly there may be no well-defined BV integration on residual fields and secondly retaining residual fields
 may be a necessary information for gluing theories from different manifolds along a common boundary.
 In the examples we consider in this paper, however, it is not necessary to work with residual fields: we will introduce them at an intermediate stage as zero modes, but we will eventually BV integrate them out.
 
 When one can BV integrate out the residual fields, the resulting state $K$
 simply satisfies the equation $\Omega K=0$. In one-dimension, the bulk is an interval 
 $[a,b]$
 and the boundary consists of two (oppositely oriented) points, so $K$ is an element of $\calH\otimes\calH^*$ and we may view it as an operator
 $U\colon\calH\to\calH$. With this identification the equation becomes $\Omega_bU-U\Omega_a=0$,
 where $\Omega_a$ and $\Omega_b$ are the $\Omega$'s corresponding to the two boundary points
 $\{a\}$ and $\{b\}$.
  
 One last remark concerns the choice of background fields. These are background choices for classical fields
 that are not fixed by the boundary conditions and the Euler--Lagrange equations. If a field $\phi$ requires
 a background field $\Phi$, we write $\phi=\Phi + \varphi$,\footnote{This is the case when the space of backgrounds is affine. In more general situations, the splitting
actually requires a choice of coordinates around each point and one has to resort to formal geometry,
see \cite{BCM} and references therein.} 
  where $\varphi$ is the fluctuation to be integrated
 (which can still be split into a residual field and a fluctuation around it).\footnote{We only allow nontrivial backgrounds for the even fields. In other words, we set the background of all odd fields to zero.
 The minimal space of residual fields is the tangent space, also in the odd directions, of the space of background fields at such a point.} 
 The BV action now depends
 on $\Phi$. Denoting by $\delta$ the de~Rham differential on the space of backgrounds, we get
 \[
 \delta\calS = \int \delta\Phi (\varphi^+,\calS).
 \]
We can express this, together with the classical master equation, more compactly as
\[
\frac12(\calS',\calS')+\delta\calS'=0,
\]
with $\calS'=\calS+\int\varphi^+\delta\Phi$. 

We can now apply the BV-BFV analysis to $\calS'$. Under the same assumptions as above
and after BV integration of the residual fields, the state $K'$
produced by the bulk now solves the equation
 \[
 \left(-\ii\hbar\delta+\Omega\right)K'=0.
 \]
If we write $K'=\sum K_i$ with $K_i$ the $i$\ndash form component on the space of backgrounds,
we get in particular $\Omega K_0=0$ and 
$\delta K_0=-\frac\ii\hbar\Omega K_1$. This shows that $K_0=K$ is $\Omega$\ndash closed and that
it changes by an $\Omega$\ndash exact term under a variation of the background field; so in cohomology
$K$ is background independent.


\subsection{One-dimensional gravity with matter.}\label{s:qonedgrav}
We start with the non extended theory in the ``abelianized'' version of Remark~\ref{r:resymmgrav}.
The first step is to define the Hilbert space of the theory that quantizes the symplectic manifold
$\calF^\de=T^*(\bbR^n\times\bbR[1])$. We choose the vertical polarization, which produces the Hilbert space
$\calH = L^2(\bbR^n)\otimes C^\infty(\bbR[1])$. The coordinates $q$ and $c$ are quantized as multiplication operators, whereas the momenta $p$ and $b$ are quantized as derivative operator. The quantization of the boundary action 
$S^\de$ is then the operator 
\[
\Omega = (\Hat H - E)\,c
\]
with
\[
\Hat H = -\frac{\hbar^2}{2m}\sum_i\frac{\de^2}{\de q_i^2} + V(q).
\]
Notice that $\Omega^2=0$ just because $c^2=0$.

An element $\psi$ of $\calH$ is a sum $\psi_0+c\psi_1$ with $\psi_0,\psi_1\in L^2(\bbR^n)$.
The cohomology of $(\calH,\Omega)$ is readily computed. In degree zero, we have $E$\ndash eigenstates
of $\hat H$: $\hat H\psi_0=E\psi_0$. In degree one, we have $L^2$ functions modulo the image
of $\Hat H_E:=\Hat H - E$:
\[
H_\Omega(\calH) = \ker\Hat H_E \oplus \coker\Hat H_E[1].
\]
The physical Hilbert space is the cohomology in degree zero $\ker\Hat H_E$, which is a quantization
of the symplectic reduction of the coisotropic submanifold $\{(p,q)\in\calF^\de: H(p,q)=E\}$ 
(when smooth).\footnote{We may also choose the polarization in which states are functions of $q$ and $b$.
Then $c$ is quantized as $-\ii\hbar\frac\de{\de b}$. A state $\phi$ will then be of the form
$\phi_{-1}b+\phi_0$ $\phi_{-1},\phi_0\in L^2(\bbR^n)$. Cohomology in degree zero is now the cokernel
of $\Hat H_E$, which in a Hilbert space can be canonically identified with the kernel of $\Hat H_E$.}

We then study the operator corresponding to bulk. We use the BV formulation explained in
Remark~\ref{r:resymmgrav} which is more convenient for the computations. For simplicity of notations
we  will remove the tildes.

To be consistent with the above polarization,
we fix the $q$ and $c$ coordinates at the boundary of the interval $[a,b]$ and leave the other fields free.
In particular, we write $C_a$ and $C_b$ to denote the values of $c$ at the boundary. We then write
$c=\Tilde C + \tau$ where $\Tilde C$ is a fixed extension in the bulk
and $\tau$ is the fluctuation.
By definition we have $\Tilde C(a)=C_a$, $\Tilde C(b)=C_b$ and $\tau(a)=\tau(b)=0$.
As explained in \cite{CMR3}, the extension $\Tilde C$ must be discontinuous with $\Tilde C(t)=0$ for
$t$ in the interior of the interval $[a,b]$. In order to avoid singularities we have to integrate by parts in order to get
rid of the terms involving derivatives of $\Tilde C$: this results in an additional boundary term: 
\[
2(g^+(b)\sqrt{g(b)}C_b-g^+(a)\sqrt{g(a)}C_a).
\]
The term involving $\tau$ is instead $2\int_a^b g^+\sqrt g\,\dot\tau\;\dd t$.

Next we rewrite the classical action \eqref{e:qg} using the change of variables
$u(t)=\int_a^t\sqrt g\;\dd t$:
\[
s(q,g) = \int_0^{u(b)} \left(
\frac12m ||q'||^2-V+E
\right)\dd u,
\]
with $q'=\frac{\dd q}{\dd u}$.
This has now the canonical form for classical mechanics in proper time $u$. 
Also notice that on the boundary we have $p=mq'+\frac{q^+c}{\sqrt g}$.
The metric $g$ is now hidden in $u(b) = \int_a^b\sqrt g\;\dd t$. 

We write $g=\gamma+\Hat g$,
where $\gamma\in\bbR_{>0}$ is a fixed constant (a background) and $\Hat g$
is the fluctuation. To keep track of the background we have to add the action an extra term:
\[
\calS'[q,q^+\!,\Hat g,\Hat g^+\!,\xi,\xi^+\!;\gamma]:=
\calS[q,q^+\!,\gamma + \Hat g,\Hat g^+\!,\xi,\xi^+] + \int_a^b \Hat g^+\delta\gamma\,\dd t,
\]
where $\delta$ is the de~Rham differential on the space $\bbR_{>0}$ of background constant metrics,
and we have denoted by $\Hat g^+=g^+$ the antifield of $\Hat g$.
We now have the following modification of the master equation:
\[
\frac12(\calS',\calS')+\delta\calS' = 0
\]
which keeps track of the dependency on the background field.

Next we have to gauge fix $\Hat g$.  First of all we have to write $\Hat g = z + \eta$, where
$z$ is a residual field (also a constant metric) to keep track of the the tangent space to background metrics
and $\eta$ denotes the fluctuation orthogonal to constant fields. As gauge fixing we choose $\eta=0$.
We also write $\Hat g^+=z^++\eta^+$ and as gauge fixing on the $(z,z^+)$\ndash space we choose $z=0$.
Finally we set $q^+=0$ and $\xi^+=0$.
The gauge fixed action (i.e., the restriction of the BV action to the gauge fixing submanifold)
is then
\begin{multline*}
\calS'_\text{gf}[q,\tau,z^+,\eta^+;C_a,C_b;\gamma]=
\int_0^{(b-a)\sqrt\gamma} \left(
\frac12m ||q'||^2-V+E
\right)\dd u+\\+
2\sqrt\gamma\int_a^b\eta^+\dot\tau\;\dd t +
\int_a^b(z^++\eta^+)\delta \gamma\,\dd t+\\+
2\sqrt\gamma[(z^++\eta^+(b))C_b-(z^++\eta^+(a))C_a].
\end{multline*}
There are now several things to notice. The first is that there are no linear terms in $\tau$, so
the linear terms in $\eta^+$ play no role in the Gaussian integral over $\tau$ and $\eta^+$. 
This then just
yields
the regularized determinant of $\frac\dd{\dd t}$ which is $1$.
The integral over $z^+$ yields the factor
\[
(b-a)\,\delta\gamma+{2\sqrt\gamma}\,(C_b-C_a).
\]
We finally have the functional integral over $q$. Note that with the gauge fixing $q^+=0$ we now have
$p=mq'$. Hence this is the usual functional integral in quantum mechanics (with potential $V-E$).
We then get that
\[
K(q_a,q_b,C_a,C_b;\gamma)= \int_{q(a)=q_a,\ q(b)=q_b,\ \tau(a)=\tau(b)=0}
\!\!\!\!\!
\!\!\!\!\!
\!\!\!\!\!
\!\!\!\!\!
\!\!\!\!\!
\!\!\!\!\!
\EE^{\frac\ii\hbar \calS'_\text{gf}[q,\tau,z^+,\eta^+;C_a,C_b;\gamma]}\;
DqD\tau D\eta^+\dd z^+
\]
is the integral kernel of the operator
\[
U = \EE^{\frac\ii\hbar\,(b-a)\,\sqrt\gamma\,(E-\Hat H)}
\left((b-a)\,\frac{\delta\gamma}{2\sqrt\gamma} + \id_C\right),
\]
where $\id_C$ is the identity on $C^\infty(\bbR[1])$ and we have rescaled by $2\sqrt\gamma$ to get the correct normalization that $U$ must be the identity operator for $b=a$.
We now explictly see that
\[
\delta U + \frac\ii\hbar (
\Omega_bU-U\Omega_a
) = 0.
\]
This means that
\[
U_0 = \EE^{\frac\ii\hbar\,(b-a)\,\sqrt\gamma\,(E-\Hat H)}
\, \id_C
\]
is $\Omega$\ndash closed and that $\delta U_0$ is $\Omega$\ndash exact, so in cohomology
$U_0$ is background independent.
%
If $\psi_0$ is a physical state (i.e., it is in the kernel of $\Hat H-E$), we get
\[
U_{0}\psi_0 = \psi_0.
\]
On the other hand, 
\[
U_{0}(\psi_1 C) = \psi_1 C + \im (E-\hat H).
\]
As expected, $U_{0}$ acts as the identity in cohomology.

\subsubsection{The extended and generalized theories}
In the extended and generalized theories we add the term $-\int_a^b E\,\dd t$ to the action.
We have three possible cases, depending on whether $E$ is fixed, it is a variable or it is a field.
We discuss the three cases in this order.

If $E$ is fixed, the discussion is as in the non extended theory with the only difference that now
$U$ gets multiplied by $\EE^{-\frac\ii\hbar (b-a)\,E}$. In particular, the operator induced in cohomology by $U_0$
becomes $\EE^{-\frac\ii\hbar (b-a)\,E}$ instead of the identity operator.
At fixed $E$ this is just a phase transformation, so it is projectively the identity operator. It is interesting however that
this operator is the correct Schr\"odinger evolution of an $E$\ndash eigenstate. We can then restore the correct time evolution if we allow for linear combinations of eigenstates of different energies. In a sense, we let $E$ vary but only after quantization.

In the second case, $E$ is already a varying parameter at the beginning. We have seen that a consequence of this is that $\xi$ (or $c$ in the other BV formulation) must vanish on the boundary and that now
$\calF^\de=T^*\bbR^n$ with coordinates $p=\frac{m\dot q}{\sqrt g}$ and $q$. 
The choice  of background and of gauge fixing is otherwise as in the non extended case. 
The gauge fixed action is then
\begin{multline*}
\calS'_\text{gf}[q,\tau,z^+,\eta^+, E;\gamma]=
\int_a^b\left(
\frac1{\sqrt \gamma}\, T(\dot q)-\sqrt\gamma\, V+ (\sqrt \gamma-1)E
\right)\dd t+\\+
2\sqrt\gamma\int_a^b\eta^+\dot\tau\;\dd t +
\int_a^b(z^++\eta^+)\delta \gamma\,\dd t.
\end{multline*}
The source term in $\eta^+$ is again irrelevant as there is no source term in $\tau$. The integral
over $\eta^+$ and $\tau$ simply produces a constant.
The integral over $z^+$ produces the factor $(b-a)\,\delta\gamma$. The integral over $E$ produces
the factor $\delta((b-a)(\sqrt\gamma -1))$. These two factors together yield
$2\delta(\gamma-1)\delta\gamma$. Integrating over $\gamma$, and getting rid of the factor $2$ to get the correct normalization,
we finally obtain
\[
K(q_a,q_b)= \int_{q(a)=q_a,\ q(b)=q_b}
\EE^{\frac\ii\hbar\int_a^b (T-V)\dd t}\;
Dq,
\]
which is the integral kernel for the usual evolution operator 
\[
U= \EE^{-\frac\ii\hbar\,(b-a)\Hat H}.
\]

Finally, in case $E$ is a field, there are  no ghosts and the BV action is equal to the classical action.
Integrating over the field $E$ we get the constraint $\sqrt g(t) = 1$ for all $t$. Inserting this into the action
yields back again the usual evolution operator.
\begin{Rem}
It is interesting that treating $E$ as a variable or as a field eventually produces the same 
quantum evolution (we already remarked in Section~\ref{s:conclusionSymplecticStructure} that they produce the same classical evolution). These are then equivalent ways of
restoring time. (Formally one might have expected this to be the case as we can approximate the field $E$ by
piecewise constants functions; given a piecewise constant $E$ we may regard evolution as the composition of the evolutions
on the intervals on which $E$ is constant, i.e., the variable case.)
We also arrive at the same result if we fix $E$ at the beginning but
allow superpositions of states (and their evolutions) for different values of $E$. In other words, we realize the evolution
operator as an integral over $E$ of the evolution operators at fixed $E$ (using the Hilbert space decomposition in 
$E$\ndash eigenspaces).
\end{Rem}


\section{AKSZ}\label{s:AKSZ}
The AKSZ formalism \cite{AKSZ} is a general construction to produce topological field theories in the BV formalism.
For a fixed integer $n$,
the target data are a finite dimensional graded supermanifold $\calY$ endowed with
a function $\sigma$ of degree $n$ and parity $n\mod2$ and with a one-form $\alpha$
of degree $n-1$ and parity $n-1\mod2$; in addition, one requires $\omega:=\dd\alpha$ to be nondegenerate
and $\{\sigma,\sigma\}=0$, where $\{\ ,\ \}$ is the graded Poisson bracket associated to $\omega$.

Then to an $n$\ndash dimensional manifold $\Sigma$ (with the same $n$ as above)
one associates a BV manifold of fields
$(\calF^\text{AKSZ}_\Sigma,\Omega^\text{AKSZ}_\Sigma,\calS^\text{AKSZ}_\Sigma)$. 
This is most easily described if one introduces Darboux coordinates $(p_i,q^i,\theta^\mu)$ on $\calY$
with $\alpha=\sum_i p_i\dd q^i + \frac12\sum_\mu\theta^\mu\dd\theta^\mu$ (note that the $\theta$ variables are odd
of degree $(n-1)/2$, so they are possibly there only if $n$ is odd). The fields are then inhomogenous
differential forms $P,Q,\Theta$ on $\Sigma$
with total degree and total parity (i.e., adding form degree and form parity to
ghost number and field parity) equal to the degree and parity of the corresponding target coordinate.

With this notation the space of fields $\calF^\text{AKSZ}_\Sigma$, which canonically is the mapping
space $\Map(T[1]\Sigma,\calY)$, has $P,Q,\Theta$ as its coordinates and one finally has
$\Omega^\text{AKSZ}_\Sigma=\int_\Sigma\left(\sum_i \delta P_i\delta Q^i + \frac12\sum_\mu\delta \Theta^\mu\delta\Theta^\mu\right)$
and 
$\calS^\text{AKSZ}_\Sigma=\int_\Sigma\left( \sum_i P_i\dd Q^i + \frac12\sum_\mu\Theta^\mu\dd\Theta^\mu
+\sigma(P,Q,\Theta)\right)$. Note that the integral selects the top degree form in each summand.

If we work on manifolds with boundary, the space of boundary fields is also a mapping space.
For $n=1$ one simply gets 
\[
(\calF^{\text{AKSZ},\de},\calS^{\text{AKSZ},\de},\alpha^{\text{AKSZ},\de})=(\calY,\sigma,\alpha).
\]

If we start with a one-dimensional theory, we can construct a para\-metrization invariant theory by using
its space of boundary fields $\calF^\de$ as target. If the original theory was already parametrization invariant, we
may expect the new one to be equivalent to it.

%

\subsection{One-dimensional gravity with matter}\label{ss-AKSZgravity}
We now apply the AKSZ formalism to the exact BFV space associated  to one-dimensional gravity with matter.
Recall from Section~\ref{s:BFVonedgrav} that $\calF^\de=T^*(\bbR^n\times\bbR[1])$,
$\alpha^\de = p\cdot\dd q+b\,\dd c$ and $S^\de =  \left(\frac{||p||^2}{2m}+V(q)-E\right)c$.
Since $\alpha^\de$ has ghost number one, this produces a one-dimensional topological field theory
with $\calF^\text{AKSZ}_I=\Map(T[1]I,\calF^\de)$. We denote the fields corresponding
to the target coordinates $p,q,b,c$ by $P,Q,B,C$, respectively, and write them as a sum of a zero-form and a one-form
with the following notation:
\begin{align*}
P &= p+q^+, & B &= -e^+ + c^+,\\
Q &= q-p^+, & C &= c-e.
\end{align*}
Note that the physical fields (i.e., ghost number zero) are the zero-forms $p,q$ and the one-form $e$, whereas
$c$ is a ghost. The BV symplectic structure is
\[
\Omega^\text{AKSZ} = \int_I (\delta p^+\delta p+\delta q^+\delta q + \delta e^+\delta e + \delta c^+\delta c)
\]
and the BV action is
\[
\calS^\text{AKSZ} = \int_I\left(
p\cdot\dd q - e^+\dd c -
\left(
\frac{||p||^2}{2m}+V(q)-E
\right)e+
\left(
\frac{q^+\cdot p}m - p^+\cdot\nabla V
\right)c
\right).
\]
{}From this we read the BRST operator Q. On the nonnegative ghost number fields we have:
\begin{align*}
Qp &= -\nabla V\,c,\\
Qq &= \frac pm\,c,\\
Qe &= -\dd c,\\
Qc &= 0.
\end{align*}
It is an immediate check that $Q^2$ is zero on all the fields and that $QS^\text{AKSZ}=0$ modulo boundary terms, where
\[
S^\text{AKSZ} = \int_I\left(
p\cdot\dd q
-
\left(
\frac{||p||^2}{2m}+V(q)-E
\right)e\right)
\]
is the classical action.

This is a first order formalism version of one-dimensional gravity with the metric field $g$ replaced by the coframe
field $e$. The abelian transformation of the coframe was first noted in \cite{BDZDVH} in the context
of the spinning particle, of which this is the nonsupersymmetric version where one can add the potential. 

Note that $\calS^\text{AKSZ}$ is well-defined for any $e$. If we however impose $e\not=0$, then
we may return to the first-order formalism integrating out the $p$ field with the BV gauge fixing $p^+=0$.
It helps writing the one-form fields in terms of the, noncanonical, one-form $\dd t$:
$q^+=-q^+_1\dd t$, 
$c^+=c^+_1\dd t$, and $e=e_1\dd t$. We obtain
\[
\Omega^\text{AKSZ}_\text{s.o.} = 
\int_I (\delta q^+_1\delta q + \delta e^+\delta e_1 + \delta c^+_1\delta c)\,\dd t
\]
and, using $\dd c = \dd t\,\dot c= -\dot c\,\dd t$,
\[
\calS^\text{AKSZ}_\text{s.o.} = \int_I\left(
\frac{T(\dot q)}{e_1}-(V(q)-E)e_1 + e^+\dot c + \frac{q^+_1\cdot \dot q c}{e_1}
\right)\,\dd t
\]
(s.o.\ stands for second order formalism).
Assuming $e_1>0$, the canonical transformation given by $g=e_1^2$ (i.e., the relation between metric and coframe)
and $g^+=\frac{e^+}{2e_1}$
yields the abelianized BV action for one-dimensional gravity with matter
of Remark~\ref{r:resymmgrav}.
\begin{Rem}[Quantization]\label{r:qAKSZfo}
We may quantize the first-order theory following the procedure explained in Section~\ref{s:qonedgrav}.
Namely, we write $I=[a,b]$,
$e=\calE+z+\eta$ (with $\calE$ the background and $z$ the residual field),
$e^+=z^++\eta^+$, $c=\Tilde C+\tau$ (with $\Tilde C$ the discontinuous extension $\Tilde C(t)=t$
for $t\in(a,b)$ of the boundary field $(C_a,C_b)$), and $c^+=\tau^+$. We consider the  extended
action $\calS^\text{AKSZ'}=\calS^\text{AKSZ}+\int_I\eta^+\delta\calE$.
With the gauge fixing
$q^+=0$, $\tau^+=0$, $\eta=0$ and $z=0$, we get
\[
U = 
\EE^{\frac\ii\hbar\,(b-a)\,\calE\,(E-\Hat H)}
\left((b-a)\,\delta\calE+ \id_C\right).
\]
Note that now we may also pick the value $\calE=0$ where $U_0$ is the identity even before passing to cohomology.
\end{Rem}

\subsection{Generalizations}\label{s:AKSZ generalizations}
Given the target manifold $\calF^\de=T^*(\bbR^n\times\bbR[1])$ with canonical one\ndash form
$\alpha^\de = p\cdot\dd q+b\,\dd c$, we may try other target actions $S^\de$. The condition that $S^\de$
be odd of degree one forces it to be of the form $S^\de=fc$ where $f$ is a function on $T^*\bbR^n$; the master equation
$\{S^\de,S^\de\}=0$ is automatically satisfied since $c^2=0$.
With the same notations as in Section~\ref{ss-AKSZgravity}, we get
\[
\calS^\text{AKSZ} = \int_I\left(
p\cdot\dd q - e^+\dd c -
fe+
\left(q^+_i\frac\de{\de p_i}f-p^{+i}\frac\de{\de q^i}f
\right)c
\right).
\]
To quantize this theory we must in general stay in the first order formalism as in Remark~\ref{r:qAKSZfo}.
One possible quantization is by perturbing around the $p\cdot\dd q$ term. The asymptotic result is given in terms
of deformation quantization. If we denote by $*$ the star product and by $\EE_*$ the corresponding
star-exponential, we get
\[
U = 
\EE_*^{-\frac\ii\hbar\,(b-a)\,\calE\,H}
\left((b-a)\,\delta\calE+ \id_C\right).
\]
In this setting the boundary action $S^\de$ is consistently quantized in terms of a star-representation.
If we take states as functions of the $q$s and consider the star product defined by the standard ordering
(all the $p$s to the right), we can write $\Omega\psi=c\,H*\psi$.

{}From the physical perspective, however, the general form of $f$ is\footnote{From now we use
Einstein's summation convention.}
\[
f(p,q)= \frac12 G^{ij}(q)(p_i-A_i(q))(p_j-A_j(q)) +W(q),
\]
where $G_{ij}(q)\dd q^i\dd q^j$ is a target metric\footnote{Note that the only thing that matters is that $G$
is nondegenerate, but otherwise it can have any signature. We present examples both with Euclidean and with Lorentzian metrics.}
(and $G^{ij}$ denotes its inverse), $A_i(q)\dd q^i$ is a target one\ndash form (the electromagnetic potential times
electric charge) and $W$ is a function.\footnote{Equivalently, one may change variables $p\mapsto p+A$
and remove $A$ from $f$ at the price of getting the one-form $(p+A)\cdot\dd q+b\,\dd c$. This is a better formulation
if $A$ is a connection but not a globally well-defined form.} 
In the previous example we had $G_{ij} = \delta_{ij}/m$, $A(q)=0$ and $W(q)=V(q)-E$.
We then get 
\begin{multline*}
\calS^\text{AKSZ} = \int_I\left(
p\cdot\dd q - e^+\dd c -
\left(
\frac12 G^{ij}(q)(p_i-A_i(q))(p_j-A_j(q)) +W(q)
\right)e+ \right.
\\ \left.
+\left(
G^{ij}(q)(p_i-A_i(q))q^+_j-
p^{+k}\frac\de{\de q^k}
\left(\frac12 G^{ij}(q)(p_i-A_i(q))(p_j-A_j(q)) +W(q)\right)
\right)c
\right).
\end{multline*}
If we assume $e$ to be nondegenerate in addition to $G$, we can integrate out $p$ with $p^+=0$
getting
\[
\calS^\text{AKSZ}_\text{s.o.} = \int_I\left(
\frac1{2e_1}G_{ij}(q)\dot q^i\dot q^j - e_1W(q)+e_1A_i(q)\dot q^i
+ e^+\dot c + \frac{q^+_1\cdot \dot q c}{e_1}
\right)\,\dd t,
\]
i.e., the BV action for the parametrization independent particle in curved space with electromagnetic field.
Proceeding as in Remark~\ref{r:qAKSZfo} we get
\[
U = 
\EE^{-\frac\ii\hbar\,(b-a)\,\calE\,\Hat H}
\left((b-a)\,\delta\calE+ \id_C\right)
\]
with $\Hat H = -\hbar^2\Delta_A + W$, where $\Delta_A$ is the covariant Laplace operator
\[
\Delta_A\phi=\frac1{\sqrt{|\det G|}} \left(\de_i+\frac\ii\hbar A_i\right)
\sqrt{|\det G|} G^{ij}\left(\de_j+\frac\ii\hbar A_j\right)\phi.
\]
Consistently, we have to quantize the boundary action as $\Omega = \hat H c$. Cohomology in degree zero is then
the kernel of $\Hat H$. Note that in the case of target Minkowski metric and $W=0$ the equation $\Hat H\psi=0$
is the Klein--Gordon equation.

\begin{Exa}
In the case $A=0$, $W=0$, and $G$ the Minkowski metric, we get $\Hat H$ the wave operator. 
Note that the second order action is classically equivalent to the Jacobi action with Minkowski metric, i.e.,
the Einstein action for the free particle.
\end{Exa}
\begin{Exa}
Choosing the target space to be $T^*(\bbR_{>0}\times\bbR)$ with base coordinates
$a>0$ and $\chi$ and choosing the target
metric $a(\dd\chi^2-\dd a^2)$, $A=0$, and $W=\frac12\left(\frac{\chi^2}a-a+\Lambda a^3\right)$ with $\Lambda$ a parameter,
we get, after integrating out the momenta, the second-order classical action
\[
S_\text{ms} =\frac12
\int_I\left(
\frac1{N}\left(
-a\dot a^2 + a\dot \chi^2
\right)-
N \left(\frac{\chi^2}a-a+\Lambda a^3\right)
\right)\dd t,
\]
with $N=e_1$. This is the minisuperspace (ms) formulation of gravity by Hartle and Hawking \cite{HH}.
Note that the potential $W$ is the non kinematical term of the Wheeler--DeWitt constraint in the minisuperspace model, whereas the kinematical term is reintepreted
as the target metric.
The explicit computations for a BV theory generated by extending Hartle and Hawking's analysis can be found in \cite{MS}. 
\end{Exa}

\subsection{Supergeneralizations}
We may add additional odd coordinates to the target. If we have a metric $G_{ij}(q)\dd q^i\dd q^j$
on $\bbR^n$, the
natural extension consists in adding $n$ odd coordinates $\theta^1$,\dots, $\theta^n$ and picking
the symplectic form $\dd (p_i\,\dd q^i + \frac12 G_{ij}(q)\theta^i\dd\theta^j)$. Moving towards supersymmetry,
we may want to double the ghost part; viz., in addition to $b$ and $c$, we now add two \emph{even}\/ coordinates $\gamma$
and $\sigma$ of degree $1$ and $-1$, respectively, and define
\[
\alpha^\de = p_i\,\dd q^i + \frac12 G_{ij}(q)\theta^i\dd\theta^j + b\,\dd c +\sigma\,\dd\gamma.
\]
The boundary action $S^\de$ is now necessarily of the form
\[
S^\de = fc + \varphi\gamma + R,
\]
where $f$ is an even function of $p,q,\theta$;
$\varphi$ is an odd function of $p,q,\theta$; and $R$ is at least linear in $b$ or $\sigma$. A necessary condition
for the master equation $\{S^\de, S^\de\}=0$ to be statisfied is that
$\{f,\varphi\}$ and $\{\varphi,\varphi\}$ are linear combinations (with functional coefficients) of $f$ and $\varphi$. Under this condition, one can always find an $R$ such that the master equation is satisfied.

The simplest choice is the case $f=\frac12\{\varphi,\varphi\}$ (note that $\{f,\varphi\}=0$ by the Jacobi identity). In this case we have
\[
S^\de = \frac12\{\varphi,\varphi\}c + \varphi\gamma + \gamma^2b.
\]
One can readily apply AKSZ.  Note that if the quantization is such that $\Hat f=\Hat\varphi^2$,
the degree zero cohomology is just the kernel of $\Hat\varphi$.

For definiteness, we consider the simplest choice
$\varphi=(p_i-A_i)\theta^i$. For simplicity we now consider the case when $G$ is constant. 
The resulting AKSZ action is that for the spinning particle in flat space, see \cite{BDZDVH,G1} and
references therein.
In particular we get
\[
f(p,q)= \frac12 G^{ij}\,(p_i-A_i(q))\,(p_j-A_j(q)) -\frac12F_{ij}(q)\theta^i\theta^j,
\]
where $F_{ij}=\de_iA_j-\de_jA_i$ is the curvature of $A$ (i.e., the electromagnetic field). 
Note that for $n=3$ the quantization of the $\theta$s is by Pauli matrices, so $\Hat\varphi^2$ is the Pauli Hamiltonian (without potential);
for $n=4$ the quantization is by gamma matrices, so $\Hat\varphi$ is
 the Dirac operator.

If we want consider a target given by a (pseudo)Riemannian manifold $(M,G)$, we have to modify the above construction in order to have globally well-defined expressions.
Namely, if we let the $\theta$s transform as vectors, the term $G_{ij}(q)\theta^i\dd\theta^j$ does not transform well. We may compensate this by letting the $p$s not transform as covectors or add an extra term. 

The most elegant way to get the correct formulae is to start from the odd tangent bundle $\Pi TM$ with base coordinates $q$ and
odd fiber coordinates $\theta$ and take its cotangent bundle $T^*\Pi TM$ with its canonical one-form: if we denote the fiber coordinates by $P$ and $\lambda$, we have
$\alpha_\text{can} = P_i\,\dd q^i + \lambda_i\,\dd\theta^i$. Note that the $\theta$s transform as vectors and the $\lambda$s as covectors, but the $P$s do not transform as covectors (the transformation has an additional term bilinear in $\lambda$, $\theta$; see below). We may get the sought after formula restricting to the symplectic submanifold
given by $\lambda_i = \frac12G_{ij}(q)\theta^i$. The resulting symplectic form is $\omega_G=\dd\alpha_G$ with
\[
\alpha_G = P_i\,\dd q^i + \frac12G_{ij}(q)\theta^i\dd\theta^j.
\]
A cute feature of this setting is that the expression
$\varphi=(P_i-A_i)\theta^i$ is a globally well-defined function. To check this we have to write down the transformation rules. Denote by $q$ and $\bar q$ two sets of coordinates on overlapping charts with
$\bar q^{\bar \imath}=\phi^{\bar \imath}(q)$. We decorate by a bar also the corresponding fiber coordinates on $\Pi TM$
and on $T^*\Pi TM$. We then get $\bar\theta^{\bar \imath}=\de_i\phi^{\bar \imath}\theta^i$,
$\lambda_j = \bar\lambda_{\bar \jmath}\de_j\phi^{\bar \jmath}$, and
\[
P_j = \de_j\phi^{\bar \imath}\bar P_{\bar \imath}+
\bar\lambda_{\bar\imath}\de_j\de_k\phi^{\bar\imath}\theta^k.
\]
Thus, since the $\theta$s are odd variables, we get
$P_j\theta^j=\bar P_{\bar\jmath}\theta^{\bar\jmath}$ and hence $\varphi(q)=\bar\varphi(\bar q)$. If we set
$\Tilde P = P-A$, we get $\varphi = \Tilde P_i\theta^i$ and the canonical one-form becomes
$\alpha_\text{can} = \Tilde P_i\,\dd q^i + A_i\,\dd q^i+\lambda_i\,\dd\theta^i$.
So far we have assumed that $A$ is a globally well-defined one-form. If $A$ is a connection, it is better
to use the last expression with $\Tilde P$ the fiber coordinate. This way, $\varphi$ is again a
globally well-defined function whereas $\alpha_\text{can}$ becomes a connection one-form and the symplectic
form $\dd\alpha_\text{can}$ is anyway globally well-defined. In the AKSZ formalism, the resulting
action is not well-defined as a function because of the term
$\int_I  A_i\,\dd q^i$, but $\EE^{\frac\ii\hbar \calS^\text{AKSZ}}$ is well-defined as a section of an appropriate line bundle (this is just the usual treatment of the charged particle).

Alternatively, one can use a connection $\Gamma$ for $TM$ in order to identify $T^*\Pi TM$
with $\Pi TM\oplus \Pi T^*M\oplus T^*M$. This simply amounts to define the momentum $p$ in the last summand by
\[
p_i = P_i + \Gamma_{ik}^l\lambda_l\theta^k.
\]
This then yields $\alpha_\text{can} = p_i\,\dd q^i - \Gamma_{ik}^l\lambda_l\theta^k\dd q^i+\lambda_i\,\dd\theta^i$
and
$\varphi = (p_i-A_i)\theta^i-\Gamma_{ik}^l\lambda_l\theta^k\theta^i$. If we choose a torsion-free
connection, then we simply get $\varphi = (p_i-A_i)\theta^i$. 

If we want to restrict to the symplectic submanifold $\lambda_i = \frac12G_{ij}(q)\theta^i$, it is convenient to use the
Levi-Civita connection. In this case, we get
\[
\alpha_G = P_i\,\dd q^i -\frac12\de_kG_{rj}\theta^r\theta^k\dd q^j+
 \frac12G_{ij}(q)\theta^i\dd\theta^j.
\]
The resulting AKSZ action is that for the spinning particle in curved space, see \cite{BDZDVH,G2} and
references therein.




\thebibliography{99}
\bibitem{AKSZ} M. Alexandrov, M. Kontsevich, A. Schwarz and O. Zaboronsky, ``The geometry of the master equation and topological quantum field theory,"
\ijmp{A12}, 1405\Ndash1430 (1997).
\bibitem{B} J. Barbour, ``The nature of time,'' \href{http://arxiv.org/abs/0903.3489}{arXiv:0903.3489}
\bibitem{BFV1} I. A. Batalin and E. S. Fradkin, ``A generalized canonical formalism and quantization of reducible gauge theories,'' Phys. Lett. {\bf B 122}, 157\Ndash164 (1983). 
\bibitem{BV81} I. A. Batalin and G. A. Vilkovisky, ``Gauge algebra and quantization,'' Phys. Lett. \textbf{B 102}, 27\Ndash31 (1981).
\bibitem{BFV2} I. A. Batalin and G. A. Vilkovisky, ``Relativistic S-matrix of dynamical systems with boson and fermion costraints,'' Phys. Lett. {\bf B 69}, 309\Ndash312 (1977).
\bibitem{BRS} C. Becchi, A. Rouet and R. Stora, ``Renormalization of
the abelian Higgs--Kibble model,'' \cmp{42},
127\Ndash162 (1975).
\bibitem{BCM}
F.~Bonechi, A.~S.~Cattaneo and P.~Mn\"ev,
``The Poisson sigma model on closed surfaces," 
JHEP \textbf{2012}, 99, pages 1\Ndash 27 (2012).
\bibitem{BDZDVH}
L. Brink, S. Deser, B. Zumino, P. Di Vecchia, and P. Howe, 
``Local supersymmetry for spinning particles,'' \pl{B 64}, 435\Ndash438 (1976).
\bibitem{CMR1} A.~S.~Cattaneo, P.~Mn\"ev and N.~Reshetikhin, ``Classical BV theories on manifolds with boundaries,'' 
\cmp{332}, 535\Ndash603 (2014).
\bibitem{CMR2} A.~S.~Cattaneo, P.~Mn\"ev and N.~Reshetikhin,
``Classical and quantum Lagrangian field theories with boundary," 
in Proceedings of the ``Corfu Summer Institute 2011 School and Workshops on Elementary Particle Physics and Gravity,"
\href{http://pos.sissa.it/archive/conferences/155/044/CORFU2011_044.pdf}{PoS(CORFU2011)044};
\href{http://arxiv.org/abs/1207.0239}{arXiv:1207.0239}
\bibitem{CMR3}  A.~S.~Cattaneo, P.~Mn\"ev and N.~Reshetikhin,
``Perturbative quantum gauge theories on manifolds with boundary,''
\href{http://arxiv.org/abs/1507.01221}{arXiv:1507.01221};
to appear in \cmp{}
\bibitem{CS1} A.~S.~Cattaneo and M. Schiavina, ``BV-BFV approach to general relativity: Einstein--Hilbert action,''
\jmp{57}, 023515 (2016), 17 pages.
\bibitem{CS2} A.~S.~Cattaneo and M. Schiavina, ``BV-BFV approach to general relativity II: Palatini--Holst action,''
in preparation.
\bibitem{G1} E. Geztler, ``The Batalin-Vilkovisky formalism of the spinning particle,"
\href{http://arxiv.org/abs/1511.02135}{arXiv:1511.02135}
\bibitem{G2} E. Geztler, ``The spinning particle with curved target,''
\href{http://arxiv.org/abs/1605.04762}{arXiv:1605.04762}
\bibitem{HH} J. B. Hartle and S. W. Hawking, ``Wave function of the Universe,''
\phr{D 28},  2960 (1983).
\bibitem{J} C. G. J. Jacobi, ``Das Princip der kleinsten Wirkung," Ch.\ 6 in
\emph{Vorlesungen \"uber Dynamik}, A. Clebsch (ed.) (1866), Reimer, Berlin;
available at \href{http://gallica.bnf.fr/ark:/12148/bpt6k902111/f53.image}{Gallica-Math}
\bibitem{KT} J. Kijowski and M Tulczyjew, \emph{A Symplectic Framework for Field Teories}, Lecture Notes in Physics \textbf{107}, Springer (1979).
\bibitem{KS} B. Kostant and S. Sternberg, ``Symplectic reduction, BRS cohomology, and infinite-dimensional Clifford algebras,'' Annals of Physics \textbf{176}, 49\Ndash113 (1987).
\bibitem{Roy} D. Roytenberg, ``AKSZ-BV formalism and courant algebroid-induced topological field theories,'' \lmp{79}, 143\Ndash159 (2007).

\bibitem{MS} M. Schiavina, \href{http://user.math.uzh.ch/cattaneo/schiavina.pdf}{\emph{BV-BFV Approach to General Relativity}}, PhD thesis, Zurich, 2015.
\bibitem{T} I. V. Tyutin,  ``Gauge invariance in field theory and statistical physics in operator formalism,''
Lebedev Institute preprint No. 39, 1975; \href{http://arxiv.org/abs/0812.0580}{arXiv:0812.0580}
\end{document}